\documentclass[]{seismica}

\usepackage{bbm}
\usepackage{soul}

\title{Correcting exponentiality test for binned earthquake magnitudes}

\author[1]{Angela Stallone
	\orcid{0000-0002-8141-017X}
	\thanks{Corresponding author: angela.stallone@ingv.it}
}
\author[2]{Ilaria Spassiani
	\orcid{0000-0003-2252-0202}
}
\affil[1]{Istituto Nazionale di Geofisica e Vulcanologia, Bologna, Italy}
\affil[2]{Istituto Nazionale di Geofisica e Vulcanologia, Rome, Italy}

\credit{Conceptualization}{Angela Stallone}
\credit{Methodology}{Angela Stallone, Ilaria Spassiani}
\credit{Software}{Angela Stallone}
\credit{Formal Analysis}{Angela Stallone, Ilaria Spassiani}
\credit{Writing - Original draft}{Angela Stallone, Ilaria Spassiani}

\begin{document}


\begin{nolinenumbers}

\makeseistitle{
	\begin{summary}{Abstract}
    Above the magnitude of completeness - the minimum threshold for which a 100\% detection rate is assumed - earthquake magnitudes are typically modeled as a continuous exponential distribution. In practice, however, earthquake catalogs report magnitudes with finite resolution, resulting in a discrete (geometric) distribution. To determine the magnitude of completeness, the Lilliefors test is commonly applied. Because this test assumes continuous data, it is standard practice to add uniform noise to binned magnitudes prior to testing exponentiality.
    Here we show analytically that uniform dithering does not recover the underlying continuous exponential distribution from its discretized (geometric) form. It instead returns a piecewise-constant residual lifetime distribution, whose deviation from the exponential model becomes detectable as catalog size or bin width increases. Through numerical experiments, we demonstrate that this deviation yields a systematic overestimation of the magnitude of completeness, with biases exceeding one magnitude unit in large, high-resolution catalogs. 
    We derive the exact noise distribution - a truncated exponential within each magnitude bin - that correctly restores the continuous exponential distribution over the whole magnitude range. Numerical tests show that this correction yields Lilliefors rejection probabilities that are consistent with the significance level across a wide range of bin widths and catalog sizes. Although illustrated for the Lilliefors test, the identified bias and the proposed correction are independent of the specific statistical test and apply generally to exponentiality testing of discretized magnitude data.
	\end{summary}
	\begin{summary}{Non-technical summary}
     Earthquake magnitudes are commonly described by a continuous exponential distribution within the range where the Gutenberg-Richter law applies. However, reported magnitudes are discretized (binned), following a geometric distribution. The Lilliefors test is often used to estimate the magnitude of completeness, and common practice is to add uniform noise to discretized magnitudes so the test can be applied. This study shows that uniform noise does not recover the underlying exponential distribution and instead introduces a bias that becomes detectable for large catalogs or coarse binning, leading to overestimation of the magnitude of completeness. We derive the correct noise that must be added, i.e., a truncated exponential within each bin, and show that it restores the continuous exponential distribution.
	\end{summary}
}
	

\section{Introduction}

Within the magnitude range above the completeness threshold where self-similarity approximately holds, the complementary cumulative frequency-magnitude distribution of earthquakes is described by the Gutenberg–Richter (GR) law \citep{gutenberg1944frequency}, a decreasing exponential model. Below this threshold, the distribution deviates from exponential behavior due to the incomplete recording of low-magnitude events, which is influenced by the quality and configuration of the seismic network (Seismic Network Density Incompleteness, SNDI) and by transient increases in noise following large earthquakes (Short-Term Aftershock Incompleteness, STAI) \cite[and references therein]{mignan2012theme}. Accurately estimating the magnitude of completeness - and thus identifying the range where the exponentiality assumption holds - is essential for any statistical analysis based on earthquake catalogs. 

The exponentiality of earthquake magnitudes should not be assumed a priori, but instead statistically verified. A common approach for testing the exponentiality of continuous data is the Lilliefors test \citep{lilliefors1969kolmogorov,clauset2009power}. Originally proposed to assess normality \citep{lilliefors1967}, the test was later adapted to exponential distributions. It is a modification of the Kolmogorov–Smirnov (KS) test \citep{massey1951kolmogorov}, specifically designed for situations in which the distribution parameters are estimated from the data and the observations are continuous. However, magnitudes reported in real seismic catalogs are discrete rather than continuous, typically provided with a resolution of one decimal place. To address this, a standard practice is to add a random variable uniformly distributed in the range $[-\Delta m/2, +\Delta m/2]$ (where $\Delta m$ is the bin width), which transforms discrete magnitudes into continuous variables. Previous studies have shown that, for $\Delta m > 0.01$, this procedure may compromise the reliability of the Lilliefors test and bias the estimates of the magnitude of completeness \citep{herrmann2021inconsistencies,spassiani2023real}.

This work was initially motivated by the observation that the Lilliefors test can return unexpectedly high estimates of the magnitude of completeness when applied to enhanced catalogs, i.e., high-resolution catalogs produced using template-matching \citep[e.g.][]{gibbons2006detection,shelly2007non,chamberlain2018eqcorrscan} or machine learning techniques \citep[e.g.][]{dokht2019seismic,zhu2019phasenet,mousavi2020earthquake}. This may come as a surprise, given that such enhanced catalogs primarily increase the number of recorded events at lower magnitudes. However, the higher rejection probability observed could, in principle, reflect true deviations from exponentiality at low magnitudes caused, for example, by improper mixing of magnitude types \citep{herrmann2021inconsistencies}, or by the greater uncertainty associated with small magnitudes \citep{pattonexploring}. While these issues could affect routine catalogs as well, the substantially larger amount of data in enhanced catalogs - usually containing more than ten times the number of events found in routine catalogs - could explain the higher rejection probabilities of the Lilliefors test, making it more sensitive to even subtle deviations from the null hypothesis of exponentiality (increased statistical power). Addressing this problem is crucial, as it can strongly affect any statistical analysis performed on high-resolution catalogs relying on the proper estimation of the magnitude of completeness \citep{herrmann2021inconsistencies,mancini2022use,stallone2025comparison}. 

An alternative explanation, however, is that the higher rejection probability actually reflects a methodological flaw in how magnitude exponentiality is tested. To rule this out, we investigate whether the common procedure of dithering binned magnitudes with uniform noise, prior to applying the Lilliefors test, systematically introduces spurious deviations from exponentiality. We address this problem both analytically and numerically, and demonstrate that dithering binned magnitudes with uniformly distributed noise fails to correctly recover the exponential distribution. Then, we derive the appropriate distribution that the added random variable must follow to ensure a correct transformation from discrete to continuous exponentially distributed values. Finally, we validate the proposed method through numerical experiments.

\subsection{Binned magnitudes}
\label{sec:binned_mag}
In real earthquake catalogs, magnitudes are reported with limited resolution. Because they take values from a countable set of observable values, their distribution is discrete rather than continuous. In probabilistic terms, this can be modeled by applying the floor function to the exponential distribution, which converts the continuous exponential distribution into its discrete analogue, i.e., the geometric distribution. More precisely, if the exponential distribution has parameter $\beta$, then the floor operation will transform it to a geometric distribution with parameter $1-e^{-\beta}$ \citep[e.g.][]{chattamvelli2020discrete,grimmett2020probability,lombardi2021normalized}. 

\cite{tinti:1987} explicitly computed the distribution of binned magnitudes for a fixed bin width. Specifically, assuming that the magnitudes are spaced by a constant bin width $\Delta m$, the variable $m_i$ representative of class $i$ can be written as
\begin{equation}
\label{eqn:geom0}
m_i = m_0 + \left(i-\frac{1}{2}\right)\Delta m,\qquad i=1,2,\dots
\end{equation} 
which determines the ordering of the bins (e.g., the first class corresponds to $m_1=m_0+\frac{\Delta m}{2}$). 
In this case, we are no longer dealing with the GR frequency-magnitude distribution for the continuous variable $m$, whose probability density function (PDF) is
\begin{equation}
\label{eqn:gr}
    p_M(m) = \beta e^{-\beta(m-m_0)}, \qquad m \ge m_0.
\end{equation}

Because the binned distribution is discrete, we must instead introduce the probability mass function (PMF) $p_i$, which can be obtained as follows:
\begin{align}
\label{eqn:pim}
    p_i &= \int_{m_i - \frac{\Delta m}{2}}^{m_i + \frac{\Delta m}{2}}\beta e^{-\beta\,(z-m_0)}dz\notag\\
    &= e^{-\beta \left(m_i - \frac{\Delta m}{2} - m_0\right)}\left[1-e^{-\beta\, \Delta m}\right]\notag\\
    &= e^{-\beta\, \Delta m(i-1)}\left[1-e^{-\beta\, \Delta m}\right]
\end{align}
where, in the last equality, we used Eq.~\eqref{eqn:geom0} to get
\begin{equation}
\label{eqn:geom00}
m_i-\frac{\Delta m}{2}-m_0 = \Delta m(i-1).
\end{equation}

Now, setting $q=1-e^{-\beta \Delta m}$, we obtain that
\begin{equation}
\label{eqn:geom}
    p_i = q\,(1-q)^{i-1}, \qquad i=1,2,\dots
\end{equation}
This proofs that earthquake magnitudes discretized according to a constant bin width $\Delta m$ follow a discrete geometric distribution with bin-dependent parameter (success probability) $q=1-e^{-\beta \Delta m}$.

Having set aside the case of non-binned magnitudes, we now analytically derive the distribution of dithered binned magnitudes. The purely theoretical case of non-binned magnitudes with added noise is addressed separately in Section S1 of the Supplemental Material.

\subsection{Uniformly dithered (binned) magnitudes}
\label{sec:uniform_dith}

As discussed in Section \ref{sec:binned_mag}, although theoretical magnitudes follow a continuous exponential distribution, in practice they are represented as binned values, consistent with the discrete resolution of real catalogs. When continuous data are required for statistical tests, such as using the Lilliefors test to estimate the magnitude of completeness, the binned magnitudes are commonly dithered by adding uniform noise.
In fact, the distribution of the random variable (r.v.) $M = M_i + Y$, where $M_i$ is discrete while $Y$ is continuous, is effectively continuous. To prove this result, we recall that a generic random variable $M$ is continuous if and only $P(M\in A)=0$ for all Borel measurable sets $A$ with zero Lebesgue measure. Now, if $M_i$ has PMF $p_i$ and $Y$ has PDF $f_Y(y)$, the distribution of $M = M_i + Y$ is given by
\begin{align*}
    f_M(m) = \sum_i p_i\,f_Y(m-i).
\end{align*}
Then, for any Lebesgue null set $A$, it holds 
\begin{align*}
    0\le P(M\in A)=\sum_i P(Y \in A-m_i \cap M_i = m_i)\le\sum_i P(Y \in A-m_i ) =0,
\end{align*}
where the last equality follows from the fact that the set $A-m_i$ is still Lebesgue null. Therefore, it follows that $P(M\in A)=0$ and $M$ is a continuous r.v.

In our specific case, using the magnitude class representation~\eqref{eqn:geom0} by \cite{tinti:1987}, we have 
\begin{align*}
&M_i = m_0 + \left(i-\frac{1}{2}\right)\Delta m,\qquad i=1,2,\dots\\
&Y\sim\mathcal{U}(a,b), 
\end{align*}
with PMF and PDF respectively given by see also
\begin{align}
\label{eqn:unif}
    &p_i = q\, (1-q)^{i-1}, \qquad q=1-e^{-\beta \Delta m}, \qquad i=1,2,\dots\notag\\
    &\notag\\
    &f_Y(y) = \begin{cases}
			     \frac{1}{b-a}, & \text{if} \quad a<y<b\\
                 0 & \text{otherwise}.
		      \end{cases}    
\end{align}

The resulting continuous distribution for $M$ can therefore be obtained in the following way:
\begin{align}
    f_M(m)&= \sum_{i=1,2,\dots}p_i \, f_Y(m-i)\notag\\
          &= \sum_{i=1,2,\dots} q\, (1-q)^{i-1}\,\frac{1}{b-a}\,\mathbbm{1}_{(a,b)}(m-i).
\end{align}
Since $f_Y(m-i)\ne0$ only if $(m-i)\in(a,b)$, or equivalently $m\in(a+i,b+i)$, the support of $f_M(m)$ is the union of the intervals $(i+a,i+b)$, for $i\in N^+$. Then
\begin{align}
    f_M(m)&= \frac{1-e^{-\beta \Delta m}}{b-a}\,\sum_{i=1,2,\dots}e^{-\beta \Delta m\,(i-1)}\, \mathbbm{1}_{(a+i,b+i)}(m),
\end{align}
where we used $q=1-e^{-\beta \Delta m}$. 
This is a piecewise constant, staircase PDF with values in the interval $(1+a,\infty)$, with decreasing steps. It precisely consists of the sum of rectangular bumps of width $b-a$, centered at every $i\in N^+$, with exponentially decaying height
\[
\frac{1-e^{-\beta \Delta m}}{b-a}\,e^{-\beta \Delta m\,(i-1)},\quad m\in\left(a+i,b+i\right),\quad i\in N^+. 
\]
If we now set 
\begin{align*}
    b &= -a = \frac{\Delta m}{2},
\end{align*} 
that is, the classical magnitude noise uniformly distributed in $(-\frac{\Delta m}{2},\frac{\Delta m}{2})$, we get
\begin{align}
    f_M(m)&= \frac{1-e^{-\beta \Delta m}}{\Delta m}\,\sum_{i=1,2,\dots}e^{-\beta \Delta m\,(i-1)}\, \mathbbm{1}_{\left(i-\frac{\Delta m}{2},i+\frac{\Delta m}{2}\right)}(m).
\end{align}
The sum above is non-zero only for the integer $i_m\in N^+$ such that $m\in\left(i_m-\frac{\Delta m}{2},i_m+\frac{\Delta m}{2}\right)$, so the final expression for $f_M(m)$ is
\begin{align}
\label{eqn:final}
    f_M(m)&= \begin{cases}
			     \frac{1-e^{-\beta \Delta m}}{\Delta m}\,e^{-\beta \Delta m\,(i_m-1)}, & \text{if}\quad m\in\left(i_m-\frac{\Delta m}{2},i_m+\frac{\Delta m}{2}\right),\quad i_M\in N^+ \\
                 0 & \text{otherwise},
		      \end{cases}
\end{align}
where $i_M$ is the largest integer $\le m+\frac{\Delta m}{2}$, and $M$ assumes values in $\left(1-\frac{\Delta m}{2},\infty\right)$. 

We can conclude that, if $M_i$ is geometrically distributed with parameter $1-e^{-\beta \Delta m}$, while $Y$ is uniformly distributed in $(-\frac{\Delta m}{2},\frac{\Delta m}{2})$, the random variable $M = M_i + Y$ follows a residual lifetime distribution, i.e., a staircase distribution of the process of holding and jump times with probability mass allocated across steps, whose PDF $f_M(m)$:
\begin{itemize}
    \item is continuous, piecewise constant on each interval $\left(i-\frac{\Delta m}{2},i+\frac{\Delta m}{2}\right)$, with support $\left(1-\frac{\Delta m}{2},\infty\right)$;
    \item consists of the infinite sum of rectangular bumps of width $\Delta m$; 
    \item the heights of the rectangular bumps geometrically decay with ratio $e^{-\beta \Delta m}$.
\end{itemize}

\subsubsection{Finite magnitudes' dataset (real case-studies)}
The calculations presented above are based on an infinite, discrete set of magnitudes, i.e., $i=1,2,\dots$. However, this assumption does not hold in real applications, where earthquake catalogs are finite and may vary significantly in quality and size, from statistically limited small datasets to more recent high-resolution records. In theoretical terms, this implies considering a truncated geometric distribution for $M_i$. Specifically, 
\begin{align*}
&M_i = m_0 + \left(i-\frac{1}{2}\right)\Delta m,\qquad i=1,2,\dots,N\\
&Y\sim\mathcal{U}\left(-\frac{\Delta m}{2},\frac{\Delta m}{2}\right), 
\end{align*}
with PMF and PDF respectively given by:
\begin{align*}
    &p_i = \frac{q}{1-(1-q)^N}\, (1-q)^{i-1}, \qquad q=1-e^{-\beta \Delta m}, \qquad i=1,2,\dots,N\\
    &\\
    &f_Y(y) = \begin{cases}
			     \frac{1}{\Delta m}, & \text{if} \quad -\frac{\Delta m}{2}<y<\frac{\Delta m}{2}\\
                 0 & \text{otherwise}.
		      \end{cases}
\end{align*}

It is important to note that here $N$ denotes the number of bins, not the total number of events in the catalog. Specifically, it represents the number of sub-intervals into which the full magnitude range is divided using a fixed, uniform bin width. 

The resulting continuous distribution for $M$ is then obtained as
\begin{align}
    f_M(m)&= \sum_{i=1}^N \frac{q}{1-(1-q)^N}\, (1-q)^{i-1}\,\frac{1}{\Delta m}\, \mathbbm{1}_{\left(i-\frac{\Delta m}{2},i+\frac{\Delta m}{2}\right)}(m)\notag\\
    &= \frac{1-e^{-\beta \Delta m}}{\Delta m}\,\frac{1}{1-e^{-\beta N\Delta m}}\,\sum_{i=1}^Ne^{-\beta \Delta m\,(i-1)}\, \mathbbm{1}_{\left(i-\frac{\Delta m}{2},i+\frac{\Delta m}{2}\right)}(m),
\end{align}
where we used $q=1-e^{-\beta \Delta m}$. This is again a piecewise constant, continuous PDF of a residual lifetime distribution, consisting of $N$  rectangular bumps of width $\Delta m$, centered at every $i=1,\dots,N$ and with exponentially decaying height (i.e., a staircase distribution). Specifically, in each interval:
\begin{align}
\label{eqn:finalN}
    f_M(m)&= \begin{cases}
			     \frac{1-e^{-\beta \Delta m}}{\Delta m}\,\frac{1}{1-e^{-\beta N\Delta m}}\,e^{-\beta \Delta m\,(i_m-1)}, & \text{if}\quad m\in\left(i_m-\frac{\Delta m}{2},i_m+\frac{\Delta m}{2}\right),\quad i_M=1,\dots,N \\
                 0 & \text{otherwise},
		      \end{cases}
\end{align}
where $i_M$ is the largest integer $\le m+\frac{\Delta m}{2}$, and $M$ assumes values in $\left(1-\frac{\Delta m}{2},N+\frac{\Delta m}{2}\right)$.

The parameter $\Delta m$ mainly changes the rectangles' width, thus controlling how well the staircase approximates the exponential function inside each bin. In this sense, it sets the ``resolution'' of the approximation. On the other hand, N controls the truncation of the support after N bins, i.e., how far the rectangles extend into the tail of the exponential distribution. By renormalizing the PDF, it rescales the whole staircase relative to the exponential.

In the following, we perform two numerical tests to isolate the impact of the bin width $\Delta m$ and the number of bins $N$ on the quality of the piecewise constant PDF derived in Eq.~\eqref{eqn:finalN}. In the first numerical experiment, we fix the maximum magnitude and vary $\Delta m$ (Figure~\ref{fig:testA}), whereas in the second we fix $\Delta m$ and vary $N$ (Figure~\ref{fig:testB}).

\begin{figure}[ht!]
\centering
\includegraphics[width=\textwidth]{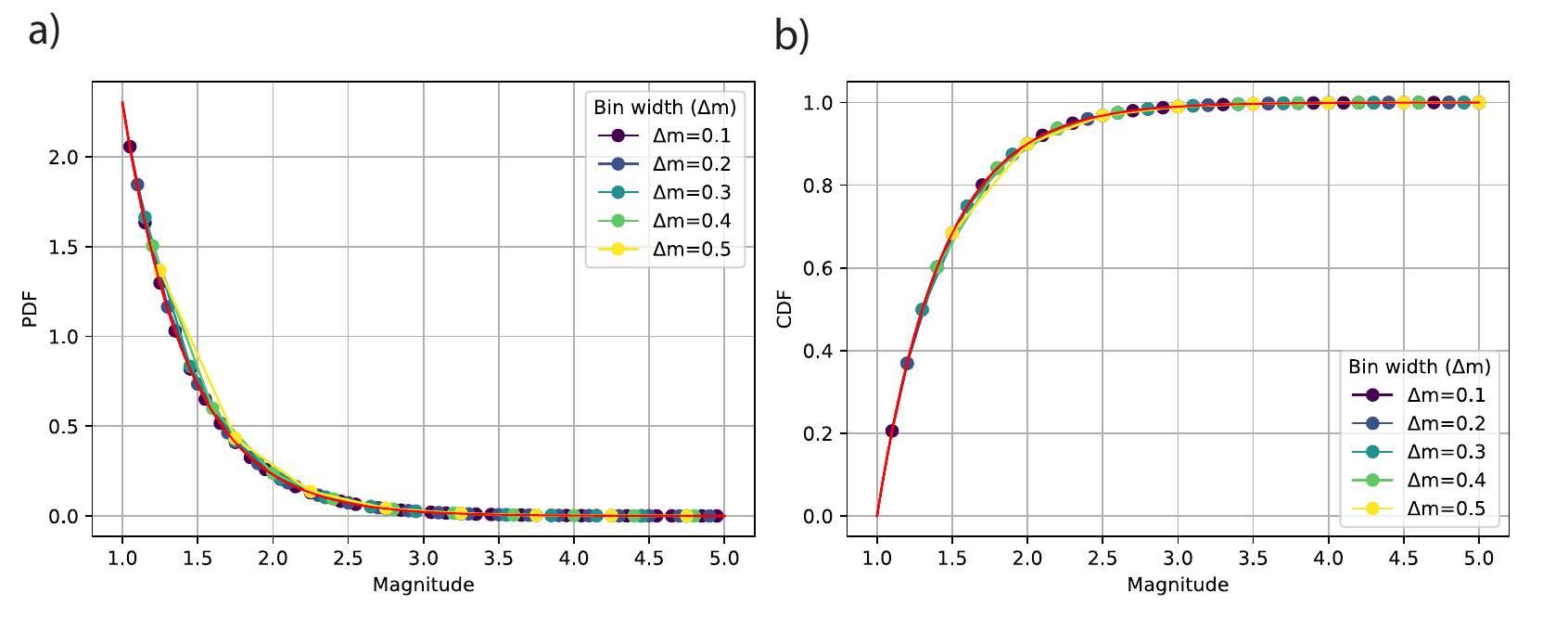}
\caption{Numerical test showing how $\Delta m$ impacts the approximation quality of the exponential distribution (solid red line). The analytical solutions of the piecewise constant staircase PDF (a) and CDF (b) are calculated for the following values of $\Delta m$: 0.1, 0.2, 0.3, 0.4, 0.5. Since the maximum magnitude is fixed ($M_{max}=5.0$), here $N$ is a dependent variable: $N = (M_{max} - M_0)/\Delta_M$, where $M_0=1.0$ is the minimum magnitude. Note: the staircase PDF is normalized by construction, i.e., the areas of the piecewise-constant rectangles of width $\Delta m$ sum to 1.
}
\label{fig:testA}
\end{figure}

\begin{figure}[ht!]
\centering
\includegraphics[width=\textwidth]{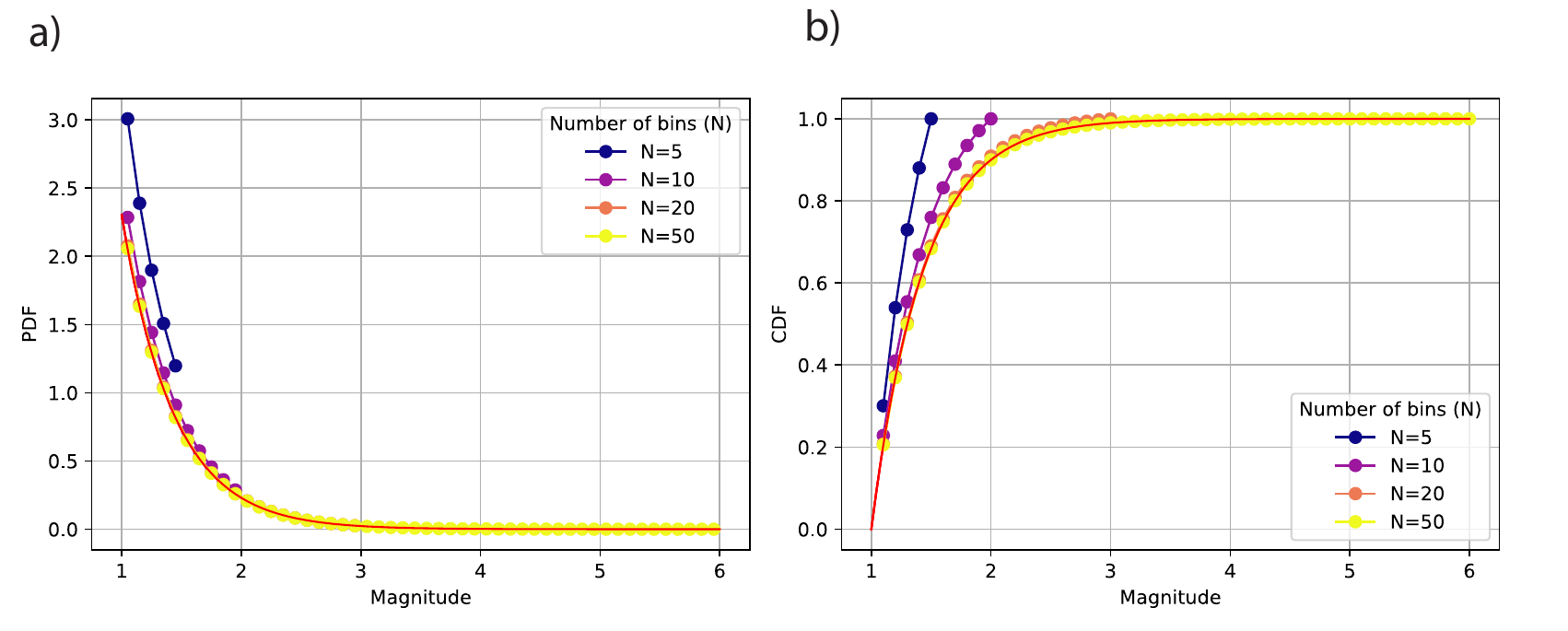}
\caption{Numerical test showing how $N$ impacts the approximation quality of the exponential distribution. Same as the previous test (see Figure~\ref{fig:testA}), but now we fix $\Delta m = 0.1$ and explore the following values for $N$: 5, 10, 20, 50.
}
\label{fig:testB}
\end{figure}
 
Shrinking the bin width $\Delta m$ forces the staircase to follow the exponential distribution more tightly, thus reducing the local approximation error. On the other hand, decreasing N affects the height of the bins, thus worsening the global fit to the exponential distribution.

\subsubsection{Numerical test: binned magnitudes dithered with uniform noise}\label{sec:num_test_uniform}

In this numerical test, we aim to answer the following question: does uniform dithering bias $M_c$ estimates?
To simulate binned magnitudes, we draw samples from the discretized GR law, i.e. the geometric distribution with parameter $p = 1 - e^{-\beta \Delta m}$.
We generate synthetic catalogs of binned magnitudes while varying only the catalog size and keeping all other parameters fixed: bin width $\Delta m = 0.1$, $b\,$-value $= 1$, and minimum magnitude $M_{\min} = 1.0$. For each catalog, we estimate the magnitude of completeness $M_c$ using the Lilliefors test ($\alpha = 0.1$, a conservative choice from a statistical point of view \citep{clauset2009power}) as implemented in the Python routine \textit{Mc-Lilliefors} \citep{marcus_herrmann_2020_4162497,herrmann2021inconsistencies}. In this idealized case, the true completeness magnitude is known ($M_c = M_{\min} = 1.0$). Therefore, any systematic bias in the estimate of $M_c$ could be indicative of a methodological flaw.

Table~\ref{Tab:Mc_Sizes} shows that the Lilliefors test correctly identifies $M_c$ (i.e., $M_c \sim M_{\min} = 1.0$) only for catalog sizes of the order of several thousand events. For larger catalogs, the estimated $M_c$ becomes systematically higher, with the error increasing as the catalog size increases. \\

\begin{table}[h]
    \centering
    \caption{Estimated magnitude of completeness $M_c$ obtained using the Lilliefors test \citep{lilliefors1969kolmogorov} for synthetic catalogs with $M_{\min}=1.0$ and varying catalog size ($\alpha = 0.1$). 
    For each catalog size, $M_c$ is calculated as the mean of the estimates obtained from 50 independent synthetic catalogs. For each synthetic catalog, 100 noise realizations are considered.}
    \label{Tab:Mc_Sizes}
    \begin{tabular}{|c|c|c|c|c|c|c|}
    \hline
    Catalog Size & 1,000 & 10,000 & 20,000 & 50,000 & 100,000 & 1,000,000  \\
    \hline
    $M_{c}$ & 1.02 & 1.04 & 1.07 & 1.34 & 1.70 & 2.65    \\
    \hline
    \end{tabular}
\end{table}

These results seem to suggest a systematic bias in statistical testing of binned magnitudes exponentiality. As the catalog size increases, the Lilliefors test rejects the null hypothesis of an exponential distribution for a progressively larger portion of the lower magnitude range. This shifts the estimated $M_c$ to larger values, even if the underlying distribution is exponential for $M \geq M_{\min}$.

To examine this issue more in detail, we set up a numerical experiment to assess how the rejection probability of the Lilliefors test varies with the bin width $\Delta m$ and catalog size. 
For various combinations of $\Delta m$ and size, we generate 1,000 synthetic catalogs of binned magnitudes. Each catalog is then dithered following the standard procedure, i.e. by adding uniformly distributed noise. Lastly, we apply the Lilliefors test. The rejection probability is defined as the proportion of simulations in which the null hypothesis of exponentially distributed magnitudes is rejected. The full procedure is described in detail in Algorithm 1 in the Supplemental Material (Section S4). Results are shown in Table~\ref{Tab:Rejection_Rates} and Figure~\ref{fig:lilliefors}.

\begin{table}[h]
    \centering
    \caption{Lilliefors test rejection probability (\%) for $\alpha$ = 0.1, based on 1,000 simulations with 100 noise realizations each, shown as a function of bin width ($\Delta m$) and catalog size.
    }
    \label{Tab:Rejection_Rates}
    \begin{tabular}{|c|ccccc|}
    \hline
    $\Delta m$ & \multicolumn{5}{c|}{Catalog size} \\
    \cline{2-6}
               & 100 & 1,000 & 10,000 & 100,000 & 1,000,000 \\
    \hline
    0.5   & 17.9  & 100 & 100 & 100 & 100 \\
    0.4  & 4.8  & 100  & 100 & 100 & 100 \\
    0.3   & 5.3   & 92.7  & 100  & 100 & 100 \\
    0.2  & 5.6   & 14.2   & 100  & 100  & 100 \\
    0.1 & 5.4   & 5.6   & 21.3  & 100  & 100  \\
    \hline
    \end{tabular}
\end{table}

The rejection probability increases with both catalog size and bin width $\Delta m$. Specifically, for a fixed catalog size, the rejection probability tends to increase as $\Delta m$ increases. With the standard bin width ($\Delta m = 0.1$), the exponential model is rejected 100\% of the time for size of order $10^5$, which corresponds to the lower range of modern high‐resolution datasets \citep[e.g.][]{ross2019searching,tan2021machine}.

These two numerical experiments show that:

\begin{enumerate}
    \item Dithering with uniform noise does not restore the original exponential distribution of magnitudes, as it introduces a systematic deviation;
    \item This deviation causes an overestimation of the magnitude of completeness $M_c$, which worsens with larger catalog sizes;
    \item For a fixed significance level, increasing catalog size and bin width leads to a higher rejection probability for the Lilliefors test, as larger samples provide more information and wider bins induce stronger deviations from the exponential model, thus increasing the power of the test.
\end{enumerate}

The apparently correct estimation of $M_c=M_{\min}$ (Table \ref{Tab:Mc_Sizes}) and the low rejection probability at low-to-intermediate catalog sizes (Table \ref{Tab:Rejection_Rates}) is an artefact caused by two opposing effects compensating each other, i.e. smoothing by dithering and the statistical power of the Lilliefors test. For catalog sizes of a few thousand events, the uniform noise smooths the geometric distribution sufficiently well that the deviations from the exponential model are too small for the Lilliefors test to detect, so the null hypothesis is not rejected. As the catalog size increases, however, the test gains statistical power and begins to detect the systematic deviations introduced by the uniform dithering, causing a growing number of rejections and an increase in the estimated $M_c$.

\begin{figure}[ht!]
\centering
\includegraphics[width=\textwidth]{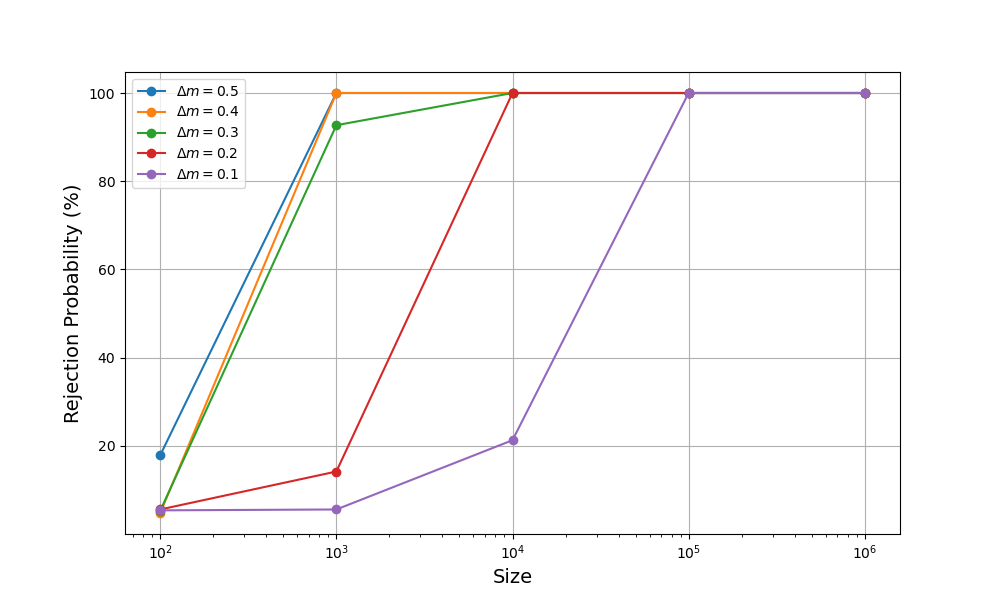}
\caption{Lilliefors test rejection probability (\%) for $\alpha$ = 0.1, computed from 1,000 simulations, each incorporating 100 noise realizations, across varying bin widths ($\Delta m$) and catalog sizes. Independent random seeds are used for each bin width and catalog size combination. See Table \ref{Tab:Rejection_Rates} for the exact probability values.}
\label{fig:lilliefors}
\end{figure}

\subsection{Exact random noise to re-obtain the exponential distribution}
Let $M_i$ be a geometric random variable with parameter $q=1-e^{-\beta\Delta m}$. We now seek to derive the exact distribution for dithering binned magnitudes, such that the exponential distribution with parameter $\beta$ is recovered for the variable $M=M_i+Y$.

Since $M_i$ creates jumps of size $\Delta m$, the final continuous exponential distribution can be seen as a continuous mixture of geometric steps, dithered by a random offset $Y\in[0,\Delta m)$. It is therefore natural to assume that the dithering variable $Y$ has support confined to this interval, i.e., $Y = 0$ outside $[0, \Delta m)$. We therefore adopt a left-edge discretization convention, where $M_i$ represents the left edge of each magnitude bin of width $\Delta m$. Note that, in Section \ref{sec:uniform_dith}, magnitudes were discretized using bin centers, following common practice (see for example the magnitude class representation~\eqref{eqn:geom0} by \citealp{tinti:1987}). While both conventions are legitimate and mathematically equivalent, the left-edge formulation is more convenient here because it avoids splitting the noise support into two symmetric intervals around the bin center (importantly, the specific discretization convention does not affect the conclusions of this work).

The convolution relating the PDFs of $M, M_i$ and $Y$ can be written as
\begin{align}
\label{eqn:fmrecover}
    f_M(m) &= \sum_{k=1}^{\infty}P(M_i = k)f_Y(m-k\Delta m), 
\end{align}
where $f_Y(m-k\Delta m)\ne0$ only if $m-k\Delta m\in[0,\Delta m)$. Then, for $m\in[k\Delta m,(k+1)\Delta m)$), the above sum simplifies to only one non-zero term. Let $y=m-k\Delta m$; since we want to recover the exponential distribution, we set $f_M(m)=\beta e^{-\beta m}$ in Eq.~\eqref{eqn:fmrecover}, obtaining
\begin{align*}
    \beta e^{-\beta(y+k\Delta m)} &= (1-e^{-\beta\Delta m})e^{-\beta\Delta m(k-1)}f_Y(y),
\end{align*}
where we used the explicit PMF in Eq.~\eqref{eqn:geom} of the geometric random variable $M_i$ (see also Eq.~\eqref{eqn:final}).
Solving the above equation for the variable $f_Y(y)$, yields
\begin{align*}
    f_Y(y) &= \frac{\beta e^{-\beta y}}{1-e^{-\beta\Delta m}}e^{-\beta\Delta m}
\end{align*}
which, after normalization $\int_0^{\Delta m}\frac{\beta e^{-\beta y}}{1-e^{-\beta\Delta m}}e^{-\beta\Delta m}dy=e^{-\beta\Delta m}$, becomes
\begin{align}
\label{eqn:finalY}
    f_Y(y) &= \frac{\beta e^{-\beta y}}{1-e^{-\beta\Delta m}}.
\end{align}

The desired $Y$ to recover the exponential distribution is therefore a truncated exponential random variable defined on the support $[0,\Delta m)$. In fact, if $X$ is a random variable with PDF and CDF respectively equal to $f_X(x)$ and $F(x)$, the truncated PDF of $X$ on the interval $(a,b]$ is obtained as $f_X(x|a<X\le b)=\frac{f_X(x)}{F(b)-F(a)}$ \cite[e.g.][]{feller1957introduction}. 
Precisely, $Y$ represents the residual waiting time within the current bin, i.e., the remainder that is obtained when sampling an exponential variable, and then dividing the result by $\Delta m$.

\subsubsection{Numerical test: binned magnitudes dithered with truncated exponential noise}

We perform two numerical experiments analogous to those described in Section \ref{sec:num_test_uniform}. 
In the first test, we generate synthetic catalogs of binned magnitudes for increasing catalog sizes, while keeping all other parameters fixed (bin width $\Delta m = 0.1$, $b\,$-value $= 1$, and $M_{\min} = 1.0$). This time, however, the binned magnitudes are dithered by adding a truncated exponential random variable supported on $[0, \Delta m)$. For each catalog size, the magnitude of completeness $M_c$ is calculated as the mean of the Lilliefors test ($\alpha = 0.1$) estimates obtained from 50 independent synthetic catalogs. For each synthetic catalog, 100 independent dithering realizations are considered in order to reduce sensitivity to individual realizations of the truncated exponential noise, following \cite{herrmann2021inconsistencies}. The resulting $p\,$-values are averaged and compared to the significance level $\alpha$. This averaged quantity should not be interpreted as a formal $p\,$-value combination, but as a way to ensure that the rejection decision does not depend on a specific noise realization.

Table~\ref{Tab:Mc_Sizes_Exp} shows that the estimated magnitude of completeness remains stable and close to the true value ($M_c \sim M_{\min} = 1.0$) across all catalog sizes, including the largest ones.

\begin{table}[h]
    \centering
    \caption{Same as Table \ref{Tab:Mc_Sizes}, but magnitudes are dithered with truncated exponential noise.}
    \label{Tab:Mc_Sizes_Exp}
    \begin{tabular}{|c|c|c|c|c|c|c|}
    \hline
    Catalog Size & 1,000 & 10,000 & 20,000 & 50,000 & 100,000 & 1,000,000  \\
    \hline
    $M_{c}$ & 1.00 & 1.01 & 1.01 & 1.02 & 1.01 & 1.03    \\
    \hline
    \end{tabular}
\end{table}

In the second numerical test, we explore how the rejection probability of the Lilliefors test varies for various combinations of $\Delta m$ and size. The procedure is summarized in Algorithm 2 in the Supplemental Material (Section S5). Results are shown in Table \ref{Tab:Rejection_Rates_Trunc} and Figure~\ref{fig:lilliefors_Trunc}.

\begin{table}[h]
\centering
\caption{Same as Table~\ref{Tab:Rejection_Rates}, but magnitudes are dithered with truncated exponential noise.}
\label{Tab:Rejection_Rates_Trunc}
\begin{tabular}{|c|ccccc|}
\hline
$\Delta m$ & \multicolumn{5}{c|}{Catalog size} \\
\cline{2-6}
           & 100 & 1,000 & 10,000 & 100,000 & 1,000,000 \\
\hline
0.5 & 0.8 & 0.5 & 0.5 & 0.8 & 1.7 \\
0.4 & 1 & 0.7 & 1.4 & 2.5 & 3.8 \\
0.3 & 2.5 & 1.9 & 3.8 & 3.8 & 6.3 \\
0.2 & 5 & 3.8 & 5.5 & 7.8 & 10.6 \\
0.1 & 5.2 & 4.3 & 7.3 & 9 & 13.7 \\
\hline
\end{tabular}
\end{table}

\begin{figure}[ht!]
\centering
\includegraphics[width=\textwidth]{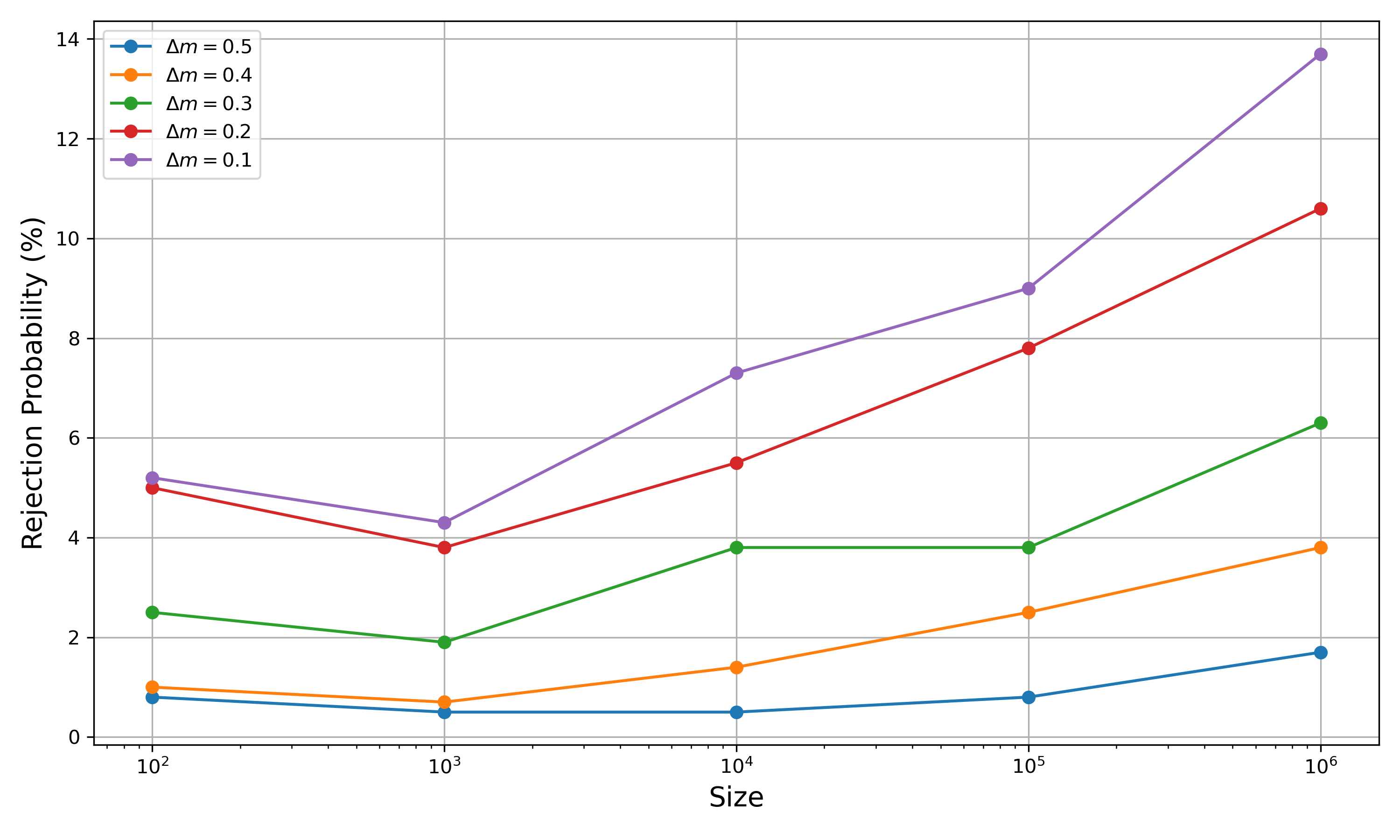}
\caption{Same as Figure~\ref{fig:lilliefors}, but magnitudes are dithered with truncated exponential noise. See also Table \ref{Tab:Rejection_Rates_Trunc}.}
\label{fig:lilliefors_Trunc}
\end{figure}

Rejection probability tends to increase with catalog size, though at a much slower rate than in the uniform–noise case. This is consistent with the statistical power of Lilliefors test increasing as the catalog size grows, meaning that even very small deviations from the exponential model are detected. These departures can arise from numerical precision effects such as tiny numerical rounding errors in computer calculations, repeated values caused by limited numerical precision, or minor inaccuracies introduced by approximate $p\,$-value calculations. Rejection probability also increases as $\Delta m$ decreases. This is plausibly related to finite–precision effects: when $\Delta m$ is small, the support of the truncated exponential noise is very narrow, and this could introduce small distortions that the test detects only when the catalog size is very large. 

Rejection probability remains below 10\% for all combinations of $\Delta m$ and $N$, except for two cases at $N=10^6$, where we observe a rejection probability of 10.6\% for $\Delta m = 0.2$ and 13.7\%  for $\Delta m = 0.1$. We interpret this behavior as resulting from numerical precision effects (see above) that are present at all sample sizes, but become more evident as the catalog size grows very large. 
Nevertheless, all rejection probabilities are far below those obtained with uniform dithering, and remain close to $\alpha=0.1$ even for $N=10^6$. This indicates that dithering magnitudes with truncated exponential noise is consistent with the assumed GR exponential model.

These conclusions remain unchanged when alternative numerical choices are adopted. In particular, using the median instead of the mean of the $N_{\mathrm{NOISE}}$ $p\,$-values (line 22 of Algorithms 1 and 2 in the Supplemental Material), or generating binned magnitudes by first sampling from the continuous GR exponential distribution and then discretizing them with the floor function, yields rejection probabilities that are statistically indistinguishable from those reported here.

\section{Discussion and conclusions} 

Validating the range over which earthquake magnitudes follow an exponential distribution is essential for any analysis that depends on a rigorous estimate of the magnitude of completeness and on parameters such as the $b\,$-value of the GR distribution. The Lilliefors test, which is generally used for this purpose, assumes continuous data. However, real catalogs report binned (discrete) magnitudes. To overcome this problem, it is common practice to dither binned magnitudes with uniform noise before applying the test. 

In this study, we show both analytically and numerically that uniform dithering does not correctly recover the underlying continuous exponential distribution. Instead, it introduces a systematic bias that remains undetected at low–to–intermediate catalog sizes due to two opposing effects neutralizing each other: the smoothing introduced by dithering and the limited statistical power of the Lilliefors test. In other words, for small and medium catalogs, the uniform noise makes the data `appear' exponential, when they actually are not. As either the bin width increases (= higher approximation error) or the catalog size grows (= more information), the Lilliefors test becomes sensitive enough to detect this deviation. The result is a systematic overestimation of the magnitude of completeness, with biases exceeding one magnitude unit in large, high-resolution catalogs. 

To address this issue, we propose replacing uniform dithering with truncated exponential dithering. We demonstrate, both analytically and numerically, that the truncated exponential noise correctly reconstructs a continuous exponential distribution without introducing artifacts into the final magnitude distribution. One could test the exponentiality hypothesis directly on binned magnitudes using methods like the maximum-likelihood estimation applied to the geometric distribution. Although feasible in principle, this is beyond the scope of this work. Our goal here is to propose a minimal modification to the established practice that completely resolves the issue identified above, while remaining simple to implement. 

The proposed correction is relevant for both standard and high-resolution catalogs, but its impact is most significant for the latter. Modern high-resolution catalogs often contain on the order of $\sim 10^6$ events, which gives the Lilliefors test a very high statistical power. When uniform dithering is applied, this increased sensitivity leads to the detection of even subtle deviations from exponentiality, resulting in a substantial overestimation of the magnitude of completeness. By contrast, the proposed truncated exponential dithering eliminates this bias and provides a statistically consistent method for completeness estimation in large modern catalogs.

\begin{acknowledgements}
We thank the Istituto Nazionale di Geofisica e Vulcanologia, Italy, grant "Progetto INGV Pianeta Dinamico” (NEMESIS, NEar real-tiME results of Physical and StatIstical Seismology for earthquakes observations, modeling, and forecasting) - code CUP D53J19000170001 - funded by Italian Ministry MIUR (“Fondo Finalizzato al rilancio degli investimenti delle amministrazioni centrali dello Stato e allo sviluppo del Paese”, legge 145/2018). 
We thank Davide Zaccagnino and Giuseppe Petrillo for their helpful insights and constructive comments.
\end{acknowledgements}

\section*{Data and code availability}

A Python implementation of the proposed truncated exponential dithering can be found on Zenodo at this link: \url{https://doi.org/10.5281/zenodo.17939386}.

\section*{Competing interests}
The authors acknowledge that there are no conflicts of interest recorded.

\bibliography{biblio.bib}
\end{nolinenumbers}
\end{document}